\documentclass[prx,a4paper,aps,twocolumn,superscriptaddress,10pt]{revtex4-1}

\usepackage{amsmath,amssymb}
\usepackage{verbatim}
\usepackage{graphicx}
\usepackage{hyperref}
\usepackage{color}
\usepackage{mathrsfs}
\usepackage[export]{adjustbox}
\usepackage{subcaption}
\usepackage{blindtext}

\newcommand {\cL}{{\cal L}}

\def\a{\alpha}
\def\b{\beta}

\def\f{\phi}
\def\g{\gamma}

\def\l{\lambda}
\def\m{\mu}
\def\n{\nu}
\def\o{\omega}

\def\L{\Lambda}

\newcommand{\be}{\begin{equation}}
\newcommand{\ee}{\end{equation}}
\newcommand{\bea}{\begin{eqnarray}}
\newcommand{\eea}{\end{eqnarray}}
\newcommand{\nn}{\nonumber}
\newcommand{\ba}{\begin{array}}
	\newcommand{\ea}{\end{array}}

\def\double #1{#1{\hbox{\kern-2pt $#1$}}}

\newcommand{\bsubeq}{\begin{subequations}}
	\newcommand{\esubeq}{\end{subequations}}

\def\ft#1#2{{\textstyle{\frac{\scriptstyle #1}{\scriptstyle #2} } }}

\begin{document}
	
	\title{Unitary Extension of Exotic Massive 3D Gravity from Bi-gravity}
	
	\author{Mehmet Ozkan}
	\email{ozkanmehm@itu.edu.tr}
	\affiliation{Department of Physics,
		Istanbul Technical University,
		Maslak 34469 Istanbul,
		Turkey}
	
	\author{Yi Pang}
	\email{Yi.Pang@maths.ox.ac.uk}
	\affiliation{Mathematical Institute, University of Oxford,
		Woodstock Road, Oxford OX2 6GG, U.K }

	\author{Utku Zorba}
	\email{zorba@itu.edu.tr}
	\affiliation{Department of Physics,
		Istanbul Technical University,
		Maslak 34469 Istanbul,
		Turkey}
	
	\date{\today}
	

	\begin{abstract}
We obtain a new 3D gravity model from two copies of parity-odd
Einstein-Cartan theories. Using Hamiltonian analysis, we demonstrate that
the only local degrees of freedom are two massive spin-2 modes. Unitarity of the
model in anti-de Sitter and Minkowski backgrounds can be satisfied for vast choices of the parameters without fine-tuning. The recent ``exotic massive 3D gravity'' model arises as a limiting case of the new model.
We also show that there exist trajectories on the parameter space of the new model which cross the boundary between unitary and non-unitary regions. At the crossing point, one massive graviton decouples resulting in a unitary model with just one bulk degree of freedom but two positive central charges at odds with the usual expectation that the critical model has at least one vanishing central charge. Given the fact that a suitable
non-relativistic version of bi-gravity has been used as an effective theory for gapped spin-2 fractional quantum Hall states, our model may have interesting applications in condensed matter physics.
		
	\end{abstract}
	
	
	\maketitle
	\allowdisplaybreaks
	
    Three-dimensional massive gravity possesses many
	surprises that are not shared by its higher dimensional cousins.
	There exists a particular combination of linear and quadratic
	curvature terms such that the theory is perturbatively unitary
	and power-counting renormalizable about the Minkowski background \cite{BHT}.
	At the linearized level, this theory reproduces the Fierz-Pauli theory propagating
	a pair of massive gravitons. It is general coordinate invariant, differing from
	the background dependent de Rham-Gabadadze-Tolley massive gravity \cite{dRGT} and thus gaining its name ``New Massive Gravity'' (NMG).
	The maximally symmetric vacua of NMG contains also the anti-de Sitter (adS) space. Unitarity in
	adS background requires in addition to the ghost-free dynamical degrees of freedom,
	the positivity of the central charges present in the asymptotic symmetry algebra. This is a requirement from
	the correspondence between quantum gravity in adS$_3$ background and two-dimensional conformal field theory (CFT) in which the central charges in the gravity model are identified with those in the dual CFT. NMG however, is not a satisfactory model to apply the adS$_3$/CFT$_2$ correspondence, as the condition under which the massive gravitons are unitary implies negative central charges.

    Another surprise in 3D gravity that many interesting models can be reformulated using frame field
    formalism as a multi-flavor Chern-Simons (CS) like theory \cite{BHMRT2}. Different from usual CS-theories, CS-like theories can bear dynamical degrees of freedom, as the structure constants disobey Jacobi identity.  The resolution of the bulk/boundary clash for NMG in adS$_3$ becomes straightforward, once adopting the frame field formalism and losing the requirement that the massive gravitons interact only through finite number of curvature terms. As a consequence, NMG is extended to the ``Zwei-Dreibein Gravity'' (ZDG) \cite{BdHMT} and further generalizations \cite{ABM} which are also parity-even. The main building blocks of ZDG are two copies of parity-even cosmological Einstein-Cartan theories glued together by a cubic potential involving the two dreibeins. Thus ZDG can be viewed as a variant of bi-gravity. By properly choosing parameters, ZDG can satisfy both criteria for a unitary adS$_3$ quantum gravity, thus serving as an attractive toy model of lower dimensional quantum gravity defined via adS$_3$/CFT$_2$ correspondence.

    Besides the more mathematically orientated goal of constructing a well-defined lower dimensional quantum gravity model, 3D spatially covariant bi-gravity has recently been utilized as a tool to build effective theories for the spin-2 gapped collective excitations observed in certain fractional quantum Hall states (FQHS) \cite{Ha,GrB,GS}. This latest application of bi-gravity in condensed matter physics is of particular interest, as it directly links two broad areas in physics, gravity, and condensed matter physics.

NMG and its two-frame fields extension ZDG are standard gravity models in the sense that the same parity property
is shared by the equations of motion and the Lagrangian. Interestingly there also exist ``exotic" 3D gravity models of which
the equations of motion are parity-even, while the actions are parity-odd. The recently proposed ``Exotic Massive Gravity" (EMG) \cite{OPT} is of just such a kind whose perturbative spectrum coincides with that of the NMG, see e.g. \cite{CGGO, GO, MOS,ATT} for recent discussions on solutions of EMG and generalizations. Similar to NMG, the full physical spectrum of
EMG contains either ghost-like massive gravitons or negative entropy Ba\~{n}ados-Teitelboim-Zanelli black holes. Simple extensions of EMG by adding parity violating terms in the Lagrangian turn out to be not sufficient
to recover unitarity \cite{OPT}. However, just as the ZDG example,
there could exist a unitary extension of EMG using two-frame fields. As we will show in later sections, such a unitary
extension (in both adS$_3$ and Minkowski background) indeed exists. Moreover, we have also found a trajectory on the parameter space interpolating between
a unitary model and a non-unitary generalized EMG living on the boundary of the parameter space. We emphasize that the previously discovered parameter flow in ZDG model \cite{PT} resides entirely in the non-unitary region and thus differs from ours per se. On the trajectory, there
is a special point separating the unitary models from the non-unitary ones. Counter-intuitively, at
the dividing point, the theory is not critical in the usual sense, namely at least one central charge vanishes \cite{LSS} with the presence of logarithmic modes \cite{GJ,CDWW}. Instead, it describes a unitary CS-like theory based on one frame field and two spin-connections.
The unitary extension of EMG retains the exotic feature of EMG as its kinetic terms are composed by two
copies of parity-odd exotic Einstein-Cartan theories \cite{Wit,TZ} in which the role of the curvature term is played by gravitational CS term.
Given the connection between bi-gravity and FQHS, it is conceivable that the new model obtained here together with the parity-even models paves the way for a fully covariant non-linear theory describing the bulk gapped collective spin-2 excitations present in a hidden sector of quantum Hall states exhibiting a parity-violating pattern.

	We begin our construction by introducing the basic fields in the frame-field formalism. We choose the same set of fields as the ZDG model, because more fields may introduce unwanted degrees of freedom.
	These are a pair of Lorentz vector-valued one forms $e^a_I$ ($I=1,2;a=0,1,2$) and a pair of Lorentz vector-valued connection one forms $\o^a_I$, out of which, one can build torsion and the Lorentz Chern-Simons actions,
	\bea
 	T^a_I (\o_J)&=& d e^a_I + \epsilon^{abc}\o_{J\,b}e_{I\,c}\,,\nn\\
	L_{\rm CS}(\o_I)&=&\ft12(\o_{I\,a} d \o_I^{a} + \ft13\epsilon^{abc} \o_{I\,a} \o_{I\,b} \o_{I\,c})\,.
	\eea
The model we start with takes the form
\be
\cL_{\rm GEZDG}=\cL_{1-}+\cL_{2-}+\cL_{12-}+\cL_{\rm v+}\,,
\label{GEZDG}
\ee
where the subscript ``GEZDG" stands for generalized exotic zwei-dreibein gravity. The ``$\pm$" in the subscript refers to the parity of the action. $\cL_{1-}$ and $\cL_{2-}$ are two exotic Einstein-Cartan Lagrangians constructed from Lorentz CS term and torsion term
\be
\cL_{I-}=\a_I L_{\rm CS}(\o_I)+ \b_I e_{I\,a}T_I^{a} (\o_I)\,,
\ee
where $\{\a_I\,,\b_I\}$ are dimensionful parameters. The two copies of exotic Einstein-Cartan theories are coupled via two other parity-odd terms
\be
\cL_{12-}=\b_3 e_{1\,a} T_{1}^a (\o_2) + \b_4 e_{2\,a} T_{2}^{a} (\o_1)\,,
\label{Lint}
\ee
where $\{\b_3\,,\b_4\}$ are new parameters. Without $\cL_{12-}$, $\{e_1\,,e_2\}$ are both invertible as independent dreibeins. As we will see from the Hamiltonian analysis below, the interaction terms (\ref{Lint}) singles out a unique invertible dreibein for the model to have the designed number of degrees of freedom. The first three terms in the Lagrangian are all parity odd. However, the odd parity implies that in the linear theory, the kinetic terms of the two massive modes have the opposite signs meaning one of them must be ghost-like. To recover unitarity, we need to add parity-even terms
\be
\cL_{\rm v+}=\epsilon^{abc}(\g_1 e_{1\,a} e_{1\,b} e_{1\,c}+\g_2 e_{2\,a}e_{2\,b}e_{2\,c}) \,,
\ee
which carry two more parameters $\{\g_1\,,\g_2\}$. This explains the origin of the name for our model. It should also be
noted that the curvature term is not present in the Lagrangian which is a special feature of the exotic model. 	
	
	We now proceed to count the degrees of freedom.
	There are in total 4 Lorentz vector-valued one-forms in \eqref{GEZDG}.
	Their temporal components are the Lagrange multipliers, thus the physical
	phase space is 24-dimensional spanned by the spatial components of the
	one-forms. As the Lagrangian is first order in time-derivative, the 24 phase
	space variables already form 12 canonical pairs. Varying the Lagrangian with respect to the 12 temporal components,
	one obtains 12 primary constraints. The integrability conditions of the first order equations of motion
   give rise to new algebraic conditions on the fields.
	We find that upon imposing $\beta_3=0$ and the invertibility of $e_2{}^a$, the integrability conditions
	lead to only two secondary constraints
	\be
	\epsilon^{ij} e_{1i\,a}\, e_{2j}^a = 0 \,,\quad \epsilon^{ij} e_{2 i\,a}\, \left( \o_{2j}^a - \o_{1j}^a \right) = 0\,,
	\label{sec}
	\ee
	and solutions to the Lagrangian multipliers associated with the temporal components of $e^a_1,\,\o^a_1 - \o^a_2$ for generic choice of the parameters. Using the procedure given in \cite{BHMRT2}, one can check
	that amongst the 14 constraints, 6 of them are first class while the rest are second class. We are then left with a $24- 12 -8 = 4$ dimensional phase space (per space point) indicating 2 degrees of freedom in the usual sense. It should also be mentioned that
	the action \eqref{GEZDG} is symmetric under $\{e_1^a\,,\o_1^a\}\leftrightarrow\{e_2^a\,,\o_2^a\}$ modulo relabelling the parameters. Thus there is a totally equivalent choice by imposing $\beta_4=0$ and invertibility of $e_1{}^a$. From now on, we will focus on the model \eqref{GEZDG} with  $\beta_3=0$ while keeping other parameters generic.
	
	The linearised spectrum about the maximally symmetric adS$_3$ vacuum should be compatible with the results from non-perturbative Hamiltonian analysis, namely, the model is free of the Boulware-Deser ghost \cite{BD, BDP} propagating only two physical massive spin-2 modes. The field equations can be readily derived from \eqref{GEZDG} from which we can read off the adS vacuum
	\be
	e_I = a_I \bar{e} \,, \quad \o_I =\bar{\o} + b_I \bar{e}\,, \label{backgroundfields}
	\ee
	provided that the parameters satisfy the relation
	\bea
	\a_1&=&-\frac{2(\b_1a_1^2 + \b_4a_2^2) }{\L + b_1^2} \,,
	\quad \a_2 = - \frac{2\b_2a_2^2 }{\L + b_2^2}\,, \nn\\
	\g_1 &=& -\frac{2 b_1 \b_1}{3 a_1} \,, \quad \g_2 = -\frac{2(b_1 \b_4 + b_2 \b_2)}{3 a_2}\,, \label{parameters}
	\eea
	where $\Lambda\equiv -1/\ell^2$ is the cosmological constant,  $\bar{e}$ and $\bar{\o} $ are the dreibein and spin connection of the unit-radius adS$_3$ metric, and $a_I$ and $b_I$ are constants. Fluctuations about the adS$_3$ vacuum are characterized by a small expansion parameter $\kappa$ as follows
	\be
	\o_I = \o + b_I e  +  \kappa v_I  \,, \quad e_I = a_I (e  + \kappa k_I)\,.
	\ee
	Diagonalizing the quadratic action (\ref{GEZDG}) about adS$_3$ vacuum is straightforward and results in
	\bea
	\cL_{\rm GEZDG}^{(2)}  &=& - \frac{K_-}{M_-} (\phi_{-\,a}D \phi_-^a - M_- \epsilon_{abc}\bar{e}^a \phi_{-}^b\phi_{-}^c)\nn\\
	&&+\frac{K_+}{M_+} (\phi_{+\,a}D \phi_+^a + M_+ \epsilon_{abc}\bar{e}^a \phi_{+}^b\phi_{+}^c)\nn \\
	&&+ a_+ (f_{+\,a}D f_+^a +\ell^{-1}\epsilon_{abc}\bar{e}^a f_{+}^bf_{+}^c)\nn\\
	&&- a_- (f_{-\,a}D f_-^a -\ell^{-1}\epsilon_{abc}\bar{e}^a f_{-}^bf_{-}^c) \,,
	\eea
	where $\{\f_{-\,a}\,,\f_{+\,a}\}$ form a pair of massive gravitons, and $\{f_{-\,a}\,,f_{+\,a}\}$ are the usual massless gravitons. The coefficients in front of the kinetic terms read
	\bea
	K_\pm&=& M_\pm(b_1 \pm M_\pm)(\ell^2M_\pm^2-1)\Delta\,,\nn\\
	a_\pm&=&  \frac{\a_1b_1\ell+\a_2b_2 \ell\pm\a_1\pm\a_2}{\ell^2} \,,\nn\\
	\Delta&=&(b_2-b_1)(M^2_--M^2_+)\b_2a^2_2\ell^{-2}\,.
	\label{diagpara}
	\eea
	 The mass eigenvalues $M_\pm$ can of course be solved in terms of the parameters in the Lagrangian. However, the expressions are not convenient for further analysis. Instead, we find that it is more handy to treat $\{M_\pm, b_1, b_2, a_1,a_2, \b_2,\ell\}$ as free parameters, recasting the original parameters $\{\a_1,\a_2,\b_1,\b_4,\g_1,\g_2\}$ in the Lagrangian as  functions of them using (\ref{parameters}) and the eigenvalue equation from which $M_\pm$ is solved. In adS$_3$ vacuum, no-tachyon and no-ghost conditions implies
	\be
	\left(\ell M_\pm \right)^2 > 1 \,,\quad  K_\pm > 0 \,.
	\label{GhostTachyon}
	\ee
	Choosing Brown-Henneaux boundary condition \cite{BH} in adS$_3$, the necessary condition for non-perturbative unitarity is the positivity of central charges appearing in the asymptotic Virasoro $\oplus$ Virasoro symmetry algebra. The central charges can be computed directly from the CS-like action using the method given in \cite{MerbisThesis}.  The final results are
	\be
	c_{\pm} = \frac{3\ell}{2G} \left( b_1 \a_1  + b_2 \a_2  \pm \frac{1}{\ell} (\a_1 + \a_2)\right) \,,
	\label{CCharge}
	\ee
	which implies that
	\be
		c_+ = \frac{3\ell}{2G} a_+\,,\quad c_- = \frac{3\ell}{2G} a_-
	\label{ctoa}
	\ee
	 In terms of the new set of parameters, $a_{\pm}$ take the form
	\bea
	a_{\pm}&=&\frac{2(b_1-b_2)(\ell M_-\pm 1)(\ell M_+\mp 1)\b_2a_2^2}{(b_2\ell \mp1)\delta}\,,\\
	\delta&=&(b_1-b_2)(1-\ell^2M_+M_-)+(M_+-M_-)(1-\ell^2b_1b_2)\,.\nn
	\label{newa}
	\eea
The tachyon-free, ghost-free and $c_\pm>0$ conditions can be simultaneously satisfied by various choices of the parameters. However, we will leave the systematic study for future work. Instead, we focus on one such region as
	\bea
	&& M_\pm \ell > 1\,,\quad M_->M_+\,,\quad b_1>b_2\,\nn\\
	&& b_1 + M_+<0\,,\quad \delta<0\,,\quad \b_2>0 \,,
	\label{UnitaryRegion}
	\eea
about which interesting physics will be discussed. $a_1$ and $a_2$ are unconstrained in \eqref{UnitaryRegion}. One representative from this unitary region takes the form
	\bea
	&&\quad M_-= 4 \,,\quad M_+= 2\,,\quad b_1=-2.5\,,\nn\\
	&&\quad b_2=-15\,,\quad \b_2=1\,,\quad a_1=a_2=1\,.
	\label{UnitaryParameters}
	\eea
	where various values are given in adS units with $\ell=1$. The spectrum analysis about Minkowski vacuum can be obtained from the adS$_3$ results by taking $\ell\to\infty$ limit. As there is no BTZ black hole in Minkowski space, the unitarity condition becomes less stringent. Only $K_\pm>0$ is required which can be easily satisfied.
	
	The GEZDG \eqref{GEZDG} model with $\b_3=0$ is in fact related to the generalized EMG by taking a scaling limit in both the fields and the parameters. Specifically, in taking the scaling limit, we redefine the fields
	\be
	e_1 = e  \,,~\o_1 = \o - \frac1{m^2} f \,,~
	e_2 = e + \frac{\l}2h \,,~\o_2 = \o  \,,
	\ee
	and the parameters
	\bea
	\a_1 &=& \frac{\n}{m^2} \,, \quad \a_2 = - 1\,,\quad\b_1 = \frac{\n}{2} - \frac{1}{\l}\,, \nn\\
	 \b_2 &=& - \frac{ \n}{2}  \,,\quad \b_4 =  \frac{1}{\l} \,,\quad \g_1 + \g_2  = \frac{\n m^2}{3\m}\,.
	 \label{FlowParameters}
	\eea
	The $\l\to 0$  limit then reproduces the generalized EMG model \cite{OPT}.
	
\begin{figure}[h!]
	\centering
	\includegraphics[scale=0.48]{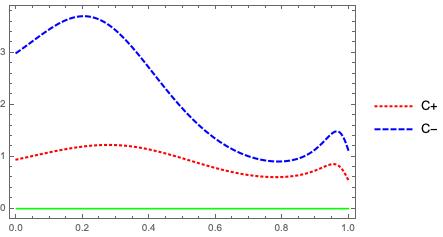}
	\caption{
	The flow of two central charges from a unitary GEZDG to a non-unitary generalized EMG model along the trajectory \eqref{parasflow}.~~~~~~~~~~~~~~~~~~~~~~~~~~~~~~~~~~~~~~~~~~~~~~~~~~}\label{cflow}
	\end{figure}

\begin{figure}[h!]
	\includegraphics[scale=0.48]{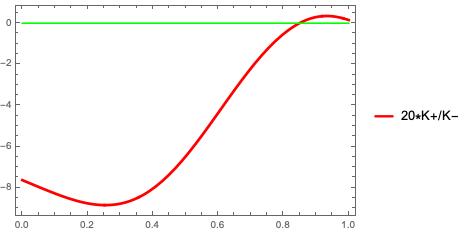}
	\caption{
		The flow of the ratio $K_+/K_-$ from a unitary GEZDG to a non-unitary generalized EMG model along the trajectory \eqref{parasflow}. At $\l=1$, $20K_+/K_-\approx0.15$.~~~~~~~~~~~
	 }\label{kflow}
	\end{figure}

\begin{figure}[h!]
	\includegraphics[scale=0.48]{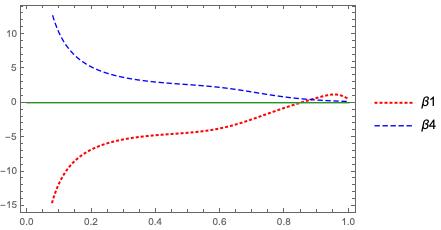}
	\caption{
	Parameters $\b_1$ and $\b_4$ flow from a unitary GEZDG to a non-unitary generalized EMG model along the trajectory \eqref{parasflow}. At $\l=1$, $\b_1\approx0.53$, $\b_4\approx0.19$.~~~~~~~}\label{bflow}
	\end{figure}

	 In fact, one can construct infinitely many trajectories connecting a point in the unitary region, such as the one given in (\ref{UnitaryRegion}), to a tachyon-free generalized EMG model. One exemplary trajectory is exhibited here on which $\ell=1$ and $\{M_\pm,b_1,b_2,a_1,a_2,\b_2\}$ are parametrized with the dependence on the flow parameter $\l$
	\bea
	M_\pm&=& \left(1-\lambda ^2\right) \mathring{M}_\pm+\lambda ^3 M_\pm^*\,,\nn\\
	b_1&=&b_1^*\lambda ^3 + \left(1-\lambda ^3\right) \Big[b_2-(\b_2 \lambda + 1) (\mathring{M}_+ - \mathring{M}_-)\Big]\,,\nn\\
	b_2&=& b_2^* \lambda ^3+\left(1-\lambda ^2\right) (\mathring{M}_+ - \mathring{M}_-+1)\,,\nn\\
	a_1&=&a_1^*\,,\nn\\
	a_2&=&a_1 \left(1-\lambda ^2\right) \left(1 - \frac{\nu \l}{4}\right)+a_2^* \lambda ^2\,,\nn\\
	\b_2&=&\b^{*}_2 \lambda ^2-   \frac{\nu}{2}\left(1-\lambda ^2\right)\,,
	\label{parasflow}
	\eea
	so that near $\l=0$, the leading behaviors of parameters above take the form \eqref{FlowParameters} up to terms linear or higher order in $\l$. When $\l=1$, the trajectory reaches the the unitary model defined by the parameters (\ref{UnitaryParameters}).
	The starred parameters represent the values in a unitary region and $\mathring{M}_\pm$ are adjustable parameters. In the $\l\to 0$ limit, we find that along the above trajectory, GEZDG reaches a generalized EMG whose defining parameters
	$\{\nu, m^2, \m\}$ are expressed in terms of $\{\mathring{M}_\pm, a_1\}$ as
	\bea
&&	 m^2=\frac{(\mathring{M}_-- 1) (\mathring{M}_++ 1)}{ a_1^2}\,, \quad \m = \frac{2 (\mathring{M}_-- 1) (\mathring{M}_++ 1)}{ a_1 (2 \mathring{M}_+ - 2 \mathring{M}_-- 1)} \,,\nn\\
	 &&	\nu=-\frac{(\mathring{M}_+ - \mathring{M}_-) (\mathring{M}_+ - \mathring{M}_-+ 2)}{ a_1^2}\,.
	\label{egmgpar}
	\eea
	We recall that the cosmological constant of the generalized EMG is given by \cite{OPT}
	\bea
	\mathring{\L}= - \Big(\nu + \frac{m^4}{\mu^2}\Big) = \frac{12  (\mathring{M}_+ - \mathring{M}_-)-1}{4 a_1^2}\,,
	\eea
	which is negative as long as we choose $\mathring{M}_- > \mathring{M}_+$. This means the limiting generalized EMG model admits adS vacuum about which the tachyon-free condition can also be satisfied if
	\bea
	m^2|\m|-m^2\sqrt{-\mathring{\L}}+|\m|\mathring{\L}>0\,.
	\eea
	Upon setting $\mathring{M}_+ = \mathring{M}_- - c$ with $c>0$ and substituting in \eqref{egmgpar}, the above condition becomes
	\bea
	- 4 c \mathring{M}_--8 c + \left( 2 c + 1\right) \sqrt{12 c+1}+4 \mathring{M}_-^2-5>0\,, \quad
	\eea
	which can be easily obeyed for large enough $\mathring{M}_-$. For instance, if $c=3$, the required condition is satisfied for any value of $\mathring{M}_-$.  With these results in hand, we may now see explicitly how the central charges $c_{\pm}$ and $K_\pm$ changes along the trajectory as $\l$ runs between 0 and 1.  In Fig.(\ref{cflow}) , Fig.(\ref{kflow}) and Fig.(\ref{bflow}), the parameters take the values given by (\ref{egmgpar}) and
	\be
	\mathring{M}_-=12\,,\quad \mathring{M}_+=9\,.
	\label{IRparameters}
	\ee
	In  Fig.(\ref{cflow}), it is evident that central charges are smooth along the trajectory and are always positive. In Fig.(\ref{kflow}), we plot $K_+$ weighted by $K_-$. The reason for this is that the magnitude of $K_\pm$ become quite large near $\l=0$ while their ratio is still modest. This ratio makes sense only when $K_-$ stays positive along the trajectory which we have also confirmed.
We also notice that $K_+$ crosses zero at $\l\sim 0.85$. At this point, $K_-$ and $c_\pm$ are still positive. This is very different from the usual intuition that the unitary and non-unitary
models are separated by critical points at which one of the central charges vanishes \cite{LSS}. This intriguing feature guides us to take a closer look at the crossing point on the trajectory.
	
From Fig.(\ref{bflow}), we see that while $\b_4$ stays positive along the entire trajectory, $\b_1$ vanishes at $\l\sim 0.85$. Meanwhile, the relation given in \eqref{parameters} implies $\g_1=0$ when $\b_1=0$. Setting $\b_1=\g_1=0$ in the GEZDG action with $\b_3=0$, we see that $e_{1\,a}$ decouples and the resulting theory is a three-flavor model given by
	\bea
	\cL_{3f}& =&\a_1 L_{\rm CS}(\o_1)+ \a_2 L_{\rm CS}(\o_2) + \b_2 e_{2\,a}T_2^a (\o_2)  \nn\\
	&& + \b_4 e_{2\,a} T^a_2 (\o_1) + \g_2 \epsilon_{abc}e_2^a e_2^b e_2^c \,.
	\label{ThreeFlavor}
	\eea
 The plots we show above indicate that it also admits unitary regions. Hamiltonian analysis reveals a single secondary constraint
	\be
	\epsilon^{ij} e_{2 i\,a}  \left( \o_{2j}^a - \o_{1j}^a \right) = 0\,,
	\ee
indicating that the model propagates a single massive graviton. We have also checked that
the new three-flavor model is inequivalent to ``Topologically Massive Gravity" \cite{DJT} or ``Minimal Massive Gravity" \cite{BHMRT}. The adS vacuum is now given by
	\be
	e_2 = a_2 \bar{e} \,, \quad \o_1 =\bar{\o} + b_1 \bar{e}\,,\quad \o_2=\bar{\o} + b_2 \bar{e}\,, \label{3fbackgroundfields}
	\ee
together with parameter relations similar to those in \eqref{parameters} upon setting $\b_1=\g_1=0$.
The quadratic action for fluctuations about the adS$_3$ background takes the form
	\bea
\cL_{3f}^{(2)} &=& - \frac{A}{M}(\phi_{\,a}D \phi^a_1 - M \epsilon_{abc}\bar{e}^{a} \phi^{b}\phi^c)\nn\\
&&+ a_+ (f_{+\,a}D f_+^a +\ell^{-1}\epsilon_{abc}\bar{e}^a f_{+}^bf_{+}^c)  \nn\\
&&- a_- (f_{-\,a}D f_-^a -\ell^{-1}\epsilon_{abc}\bar{e}^a f_{-}^bf_{-}^c) \,,
\eea
where $a_\pm$ are  still related to central charges via (\ref{ctoa}). In terms of the new parameters \{$M, b_1, b_2,a_2, \b_2,\ell$\}, $a_\pm$ and $A$ are given as
\bea
a_\pm &=& \frac{2 a_2^2 \ell \b_2 \left(b_2 - b_1 \right) \left(M \ell \pm 1\right)}{\left(b_2 \ell \mp 1\right) \left(b_1 \ell \mp 1\right) \left(b_2 + M\right)} \,,\nn\\
A &=& M \left(b_1 - b_2\right) \left(b_1 + M\right) \left(\ell^2 M^2 - 1 \right) \,.
\eea
The bulk/boundary unitarity is achieved when
\be
A > 0 \,, \quad a_\pm > 0 \,,
\ee
which can be satisfied for various choices of parameters. In adS units, one example is given for the choice of parameters below
\be
b_1 = -3\,, \quad b_2 = -12\,, \quad M = 5 \,, \quad \b_2 = a_2 = 1 \,.
\ee

  In this letter, we report a new unitary four-flavor CS-like model obtained
	from merging together two copies of parity-odd exotic Einstein-Cartan theories. The generalized EMG model appears as a limiting case of the new model. Regarding the connection between bi-gravity and effective theory for the gapped spin-2 FQHS, the exotic nature of the new model may describe certain novel phenomena in fractional quantum Hall effects once a non-relativistic limit is properly taken. From the effective field theory point of view, it is crucial to understand the boundary on the parameter space separating the unitary models from the non-unitary ones. Therefore it should be interesting to carry out a systematic study of all unitary four-flavor CS-like theories with a bi-gravity origin. Besides the standard and exotic four-flavor CS-like theories, there exists a third kind of mixed nature by coupling a parity-even theory to a parity-odd one. Unitary models of this type have not been investigated up to date. Finally, incorporating supersymmetry in the CS-like theory is also interesting. It has been proposed that global supersymmetry may emerge in certain condensed matter systems, e.g. \cite{PL}. If in certain FQHS emergent local supersymmetry can be realized, its effective theory must be a supersymmetric bi-gravity model, thereby revitalizing supergravity in a new arena.

\noindent\textbf{Acknowledgements}:	We thank P.~K.~Townsend for stimulating discussions and D.~Grumiller for helpful correspondence. The work of M.O. is supported in part by TUBITAK grant 118F091. The work of Y.P. is supported by a Newton International Fellowship NF170385 of the UK Royal Society. U.Z. is supported in parts by Istanbul Technical University Research Fund under grant number TDK-2018-41133.

\end{document}